\documentclass[twocolumn,longbibliography,superscriptaddress]{revtex4-2}
\usepackage{graphicx}
\usepackage{dcolumn}
\usepackage{bm}
\usepackage{color}
\usepackage{braket}
\usepackage{amsmath}
\usepackage{physics}
\usepackage[caption=false]{subfig} 

\newcommand{\Nbitstring}{{\{0, 1\}^{N_q}}}
\newcommand{\device}[1]{$\mathsf{ibm\_#1}$} 

\begin{document}

\title{Quantum Computation of a Quasiparticle Band Structure with the Quantum-Selected Configuration Interaction}

\author{Takahiro Ohgoe}
\affiliation{Panasonic Holdings Corporation, 1006 Kadoma, Kadoma City, Osaka 571-8508, Japan}

\author{Hokuto Iwakiri}
\affiliation{QunaSys Inc., Aqua Hakusan Building 9F, 1-13-7 Hakusan, Bunkyo, Tokyo 113-0001, Japan}

\author{Kazuhide Ichikawa}
\affiliation{Panasonic Holdings Corporation, 1006 Kadoma, Kadoma City, Osaka 571-8508, Japan}

\author{Sho Koh}
\affiliation{QunaSys Inc., Aqua Hakusan Building 9F, 1-13-7 Hakusan, Bunkyo, Tokyo 113-0001, Japan}

\author{Masaya Kohda}
\affiliation{QunaSys Inc., Aqua Hakusan Building 9F, 1-13-7 Hakusan, Bunkyo, Tokyo 113-0001, Japan}

\date{\today}

\begin{abstract}
Quasiparticle band structures are fundamental for understanding strongly correlated electron systems. While solving these structures accurately on classical computers is challenging, quantum computing offers a promising alternative. Specifically, the quantum subspace expansion (QSE) method, combined with the variational quantum eigensolver (VQE), provides a quantum algorithm for calculating quasiparticle band structures. However, optimizing the variational parameters in VQE becomes increasingly difficult as the system size grows, due to device noise, statistical noise, and the barren plateau problem. To address these challenges, we propose a hybrid approach that combines QSE with the quantum-selected configuration interaction (QSCI) method for calculating quasiparticle band structures. QSCI may leverage the VQE ansatz as an input state but, unlike the standard VQE, it does not require full optimization of the variational parameters, making it more scalable for larger quantum systems. Based on this approach, we demonstrate the quantum computation of the quasiparticle band structure of a silicon using 16 qubits on an IBM quantum processor.
\end{abstract}

\maketitle

\section{\label{sec:intro}Introduction}
Accurately describing quasiparticle band structures is crucial for understanding strongly correlated electron systems, as these structures provide key insights into the fundamental properties of materials. Despite the advancements in computational techniques, solving these structures precisely remains a significant challenge for classical computers. Traditional methods such as density functional theory (DFT) have achieved substantial success in simulating materials~\cite{hohenberg1964,kohn1965}, but they fall short in capturing strong electron correlations and systematically underestimate the band gaps in semiconductors and insulators. More sophisticated approaches, including many-body perturbation theories like the GW approximation~\cite{hybertsen1985}, partially address these limitations but still struggle with strongly correlated systems. Wave-function-based methods, originally developed in quantum chemistry, have shown promise for tackling these issues~\cite{mcclain2017,dittmer2019,wang2020,gallo2021}, yet their computational costs render them impractical for large-scale solid-state systems.

Quantum computing has emerged as a powerful alternative, offering the potential to efficiently and accurately simulate quantum systems. Two prominent quantum algorithms for electronic structure calculations are the quantum phase estimation (QPE)~\cite{kitaev1995quantum,cleve1998,guzik2005} and the variational quantum eigensolver (VQE)~\cite{peruzzo2014}. QPE can theoretically provide exponential speedup in exact full configuration interaction (FCI) calculations. However, its reliance on deep quantum circuits makes it highly susceptible to noise, limiting its applicability on current noisy intermediate-scale quantum (NISQ) devices~\cite{preskill2018}. On the other hand, VQE uses shallow quantum circuits, making it more suitable for near-term applications and enabling successful demonstrations in small-molecule simulations~\cite{peruzzo2014,kandala2017,kandala2019}. This adaptability has spurred significant interest in extending VQE to more complex systems, including periodic materials.

Quasiparticle band structures have recently been studied using VQE combined with quantum subspace expansion (QSE) methods~\cite{mcclean2017,fan2021,yoshioka2022}\footnote{
From a different perspective, there is a proposal~\cite{kanno2021qml} to employ quantum machine learning for band-structure calculations, with a demonstration~\cite{sureshbabu2021implementation} on a quantum device.
}. These methods, which are inspired by the equation-of-motion approach, enable the computation of excited states and quasiparticle band structures. So far, most of experimental demonstrations of band structures on real quantum devices have been limited to tight-binding Hamiltonians~\cite{cerasoli2020,sherbert2021,zhang2024}, which neglect two-body interactions critical for describing electron correlations. However, incorporating two-body terms is essential for practical first-principles calculations especially for strongly correlated systems. In our previous study~\cite{ohgoe2024}, we have experimentally demonstrated the calculation of quasiparticle band structure of a silicon based on the Hamiltonian including the two-body terms. However, the number of qubits utilized on a real quantum device was limited to 2, which results in the calculation of only two bands. Increasing the number of qubits becomes difficult, because optimizing the variational parameters in VQE becomes severer as the system size grows, due to device noise, statistical noise~\cite{gonthier2022measurements}, and the barren plateau problem~\cite{mcclean2018barren}.

\begin{figure*}
    \includegraphics[width=\textwidth]{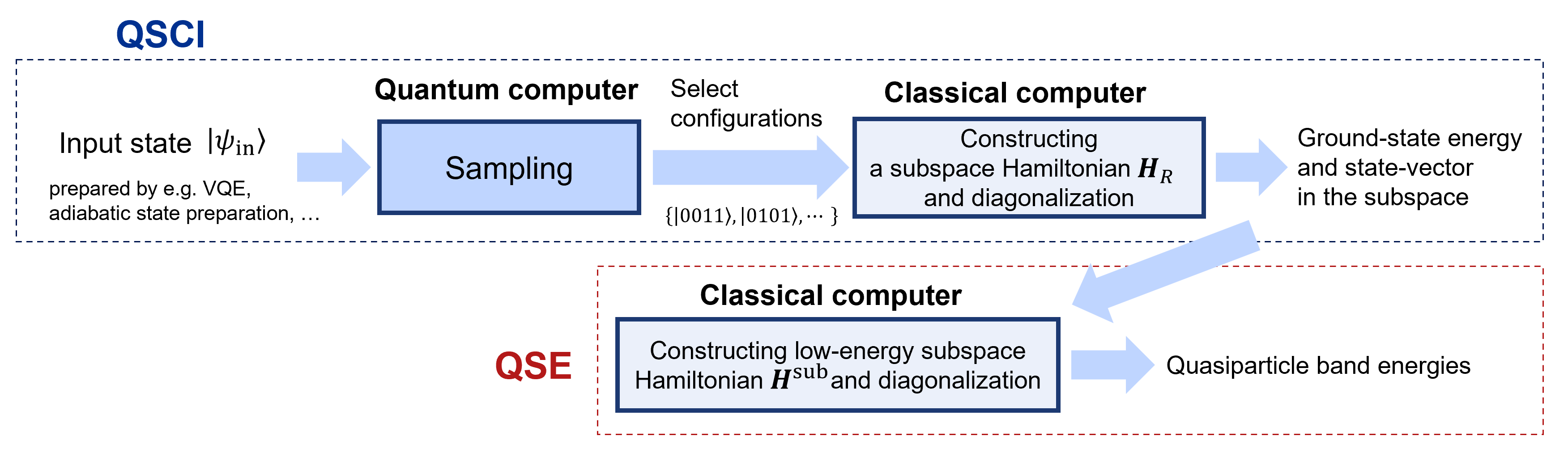}
    \caption{(Color online) Schematic description of our approach to calculate the quasiparticle band structure in this paper. It is based on QSCI for the ground-state calculation, combined with QSE to find quasiparticle bands.}
    \label{fig:flowchart}
\end{figure*}

In this paper, we propose a hybrid approach that combines QSE with the quantum-selected configuration interaction (QSCI) method~\cite{kanno2023}. Here, QSCI is introduced as an alternative to VQE for ground-state calculations. It utilizes quantum computers to sample relevant electron configurations and classical computers for subspace Hamiltonian construction and diagonalization.
Since the proposal in Ref.~\cite{kanno2023}, QSCI has been actively studied~\cite{nakagawa2024,kaliakin2024accurate,Barison_2025,liepuoniute2024quantum,shajan2024towards,sugisaki2024,mikkelsen2024quantum,reinholdt2025exposing,yu2025quantum,kaliakin2025implicit,yoshida2025auxiliary,danilov2025enhancing,barroca2025surfacereactionsimulationsbattery,duriez2025computingbandgapsperiodic}, including a large-scale experimental demonstration~\cite{robledo2024chemistry}.
QSCI may leverage the VQE ansatz as an input state but, unlike the standard VQE, it does not require full optimization of the variational parameters, improving scalability for larger quantum systems. To demonstrate the feasibility of this approach, we calculate the quasiparticle band structure of a silicon crystal using an IBM quantum processor. Our results highlight the potential of hybrid quantum-classical algorithms for advancing the study of strongly correlated electron systems.

\section{Hamiltonian representation}
We consider the second-quantized representation of {\it ab\ initio} Hamiltonians of periodic systems, which is expressed as follows:
\begin{eqnarray}
  {\hat H} & = & \sum_{pq} \sum_{\mathbf k} t_{pq}^{\mathbf k} {\hat c}_{p{\mathbf k}}^{\dagger}  {\hat c}_{q{\mathbf k}}^{\vphantom{\dagger}}  \nonumber  \\
             &    &  +  \sum_{pqrs}  \sum_{{\mathbf k}_p {\mathbf k}_q {\mathbf k}_r {\mathbf k}_s}^{\prime}  v_{pqrs}^{{\mathbf k}_p {\mathbf k}_q {\mathbf k}_r {\mathbf k}_s} {\hat c}_{p{\mathbf k}_{p}}^{\dagger} {\hat c}_{q{\mathbf k}_{q}}^{\dagger} {\hat c}_{r{\mathbf k}_{r}}^{\vphantom{\dagger}} {\hat c}_{s{\mathbf k}_{s}}^{\vphantom{\dagger}}. \label{eq:ham} 
\end{eqnarray}
Here, ${\hat c}_{p{\mathbf k}}^{\dagger}$ (${\hat c}_{p{\mathbf k}}^{\vphantom{\dagger}} $) is the fermionic creation (annihilation) operator on the $p$th crystalline orbital (CO) with the crystal momentum ${\mathbf k}$, which is obtained from Hartree-Fock (HF) calculations. The complex coefficients $t_{pq}^{\mathbf k}$ and $v_{pqrs}^{{\mathbf k}_p {\mathbf k}_q {\mathbf k}_r {\mathbf k}_s}$ represent one-body and two-body integrals between COs, respectively. The prime notation on the summation symbol indicates that the momentum conservation law, ${\mathbf k}_p + {\mathbf k}_q - {\mathbf k}_r - {\mathbf k}_s = {\mathbf G}$ must be satisfied, where ${\mathbf G}$ is a reciprocal lattice vector of the unit cell.

For quantum computations, the fermionic Hamiltonians Eq.~(\ref{eq:ham}) are mapped into the qubit Hamiltonians. In the present work, we employ the Jordan-Wigner transformation~\cite{jordan1928} as the mapping technique.

\section{\label{sec:method}Quantum algorithms}

To calculate quasiparticle band structures, we take a hybrid quantum-classical approach.
In our previous study~\cite{ohgoe2024}, we employed VQE combined with QSE.
It is based on the formulation of Ref.~\cite{yoshioka2022}, where VQE is used to calculate the ground-state wave function and its energy while QSE to find quasiparticle excitations on the VQE state, which we identify as the quasiparticle bands.

In this study, we adopt QSCI for the ground-state calculation, combined with QSE to find the quasiparticle band structure.
Figure~\ref{fig:flowchart} illustrates an overview of our approach.
In this setup, the QSE part is performed by classical computation with a compact representation of the ground-state wave function output by QSCI, while QSCI utilizes quantum computers.
We expect this QSE-QSCI approach is applicable to larger system sizes than the original QSE-VQE setup.
The following subsections provide a detailed explanation of our approach that integrates QSCI and QSE, describing each method in turn.

\subsection{QSCI for Ground States}

QSCI~\cite{kanno2023} is a class of hybrid quantum-classical algorithms to find energy eigenvalues and eigenstates for many-body systems. 
Taking an approximate ground state of electronic structure problem as input, QSCI can output a better approximation of the ground-state energy with a compact representation of the corresponding wave function. It is a subspace diagonalization method, where quantum computers are harnessed to select the electron configurations spanning the subspace. The algorithm is briefly summarized below.

For an approximate ground state $\ket{\psi_{\rm in}}$, taken as an input of the algorithm, we perform the sampling task on quantum computers, where the preparation of the input state $\ket{\psi_{\rm in}}$ and measurement in the computational basis are repeated by $N_{\rm shot}$ times. Each measurement yields a computational basis state $\ket{x}$, where $x$ is an $N_q$ bit string $x\in\Nbitstring$ with $N_q$ the number of qubits. Then, we pick up $R$ most frequent computational basis states in the sampling outcomes to form a subset of the whole computational basis:
\begin{align}
\mathcal{S}_R = \{ \ket{x} | x\in\Nbitstring, R~{\rm most~frequent} \}.
\label{eq:subset}
\end{align}
Note that each $\ket{x}$ corresponds to a Slater determinant, or an electron configuration, under some fermion-qubit mapping. Hence, $\mathcal{S}_R$ defines a subspace of the entire Fock space that is tailored for the ground state.

We then construct a subspace Hamiltonian $\bm{H}_R$ that is an $R\times R$ matrix defined by $(\bm{H}_R)_{xy}= \mel{x}{\hat{H}}{y}$ for $\ket{x}, \ket{y} \in \mathcal{S}_R$, and solve the eigenvalue problem $\bm{H}_R\bm{c} = E_R\bm{c}$, where $\bm{c}$ is an eigenvector with the eigenvalue $E_R$. The ground-state energy and a compact representation of its wave function are given as the lowest eigenvalue $E_R$ and corresponding $\bm{c}$, respectively.
Accordingly, we define the output state $\ket{\psi_{\rm out}}$, which would better approximate the ground state, by
\begin{align}
    \ket{\psi_{\rm out}} =\sum_{\ket{x}\in \mathcal{S}_R} c_x \ket{x}.
    \label{eq:output-state}
\end{align}
Here, $c_x$ is an element of $\bm{c}$, with the normalization condition $\bm{c}^\dagger \bm{c}=1$ assumed.

This algorithm utilizes quantum computers exclusively to identify electron configurations relevant to an accurate description of the ground state. Meanwhile, the construction of the subspace Hamiltonian and its diagonalization are both performed using classical computation.
In this way, QSCI utilizes quantum computers only during the intermediate step of the overall algorithm.
Note that measuring energy expectation values on quantum computers is highly susceptible to device noise and statistical error, while QSCI avoids such measurements, leading to its robustness against noise.
We can further mitigate the noise effects by post-selecting the sampling results. Specifically, a sampling outcome $\ket{x}$ is discarded if it is not compatible with conserved quantities such as the particle number and the spin of electrons.
In this case, the subspace $\mathcal{S}_R$ of Eq.~\eqref{eq:subset} is redefined by $R$ most frequent configurations that satisfy the post-selection criteria.

Now we consider how to prepare the input state for QSCI. In this study, we use VQE~\cite{peruzzo2014} for the state preparation. 
VQE is a variational method with a parametrized ansatz for electron wave function, represented as $|\psi ({\bm \theta}) \rangle = {\hat U} ({\bm \theta}) \ket{\psi_0}$.
Here, ${\bm \theta}$ is a set of variational parameters to be optimized to minimize the energy expectation value, and ${\hat U} ({\bm \theta})$ is a unitary operator implemented by a quantum circuit acting on some initialized state $\ket{\psi_0}$.
In the current QSCI setup, VQE does not require full optimization of the parameters as discussed in Ref.~\cite{kanno2023}. A moderately optimized ansatz state can still serve as a good input state so long as it contains dominant electron configurations to describe the ground state, even if the weights of configurations in the wave function are not fully optimized.

\subsection{QSE for Qausiparticle Band Structure}

We then apply QSE~\cite{mcclean2017} to the output state of QSCI, $\ket{\psi_{{\rm out}}}$ in Eq.~\eqref{eq:output-state}, which approximates the ground state.
With an excitation operator ${\hat O}_{i}$, we construct a low-energy subspace Hamiltonian ${\bm H}^{\rm sub}$ that is defined by $H_{ij}^{\rm sub} = \langle \psi_{\rm out} | {\hat O}_{i}^{\dagger} {\hat H} {\hat O}_{j}^{\vphantom{\dagger}} | \psi_{\rm out} \rangle$ in the matrix representation. Then we obtain the spectrum of energy eigenvalues (quasiparticle bands) by solving the generalized eigenvalue problem ${\bm H}^{\rm sub} {\bm C} = {\bm S}^{\rm sub} {\bm C} {\bm E}$. Here, ${\bm C}$ is the matrix consisting of the eigenvectors and ${\bm E}$ is the diagonal matrix whose elements are eigenvalues. In addition, the matrix element of ${\bm S}^{\rm sub}$ is defined by $S_{ij}^{\rm sub} = \langle \psi_{\rm out} | {\hat O}_{i}^{\dagger} {\hat O}_{j}^{\vphantom{\dagger}} | \psi_{\rm out} \rangle$.
As the excitation operator ${\hat O}_{l {\mathbf k}}^{\vphantom{\dagger}}$, we choose ${\hat c}_{l {\mathbf k}}^{\vphantom{\dagger}}$ for obtaining the valence bands at each ${\mathbf k}$, where $l$ runs over the occupied orbitals. Similarly, ${\hat c}_{l {\mathbf k}}^{\dagger}$ is used for the conduction bands, where $l$ runs over the unoccupied orbitals.

The matrix elements $H_{ij}^{\rm sub}$ and $S_{ij}^{\rm sub}$ are supposed to be evaluated using quantum devices in the original QSE-VQE setting~\cite{yoshioka2022}.
On the other hand, in the current QSE-QSCI setup, we calculate them by classical computation with a compact representation of the ground-state wave function $\ket{\psi_{{\rm out}}}$, obtained by QSCI.
Overall, we use classical computation for both QSCI and QSE to evaluate the matrix elements and to solve the (generalized) eigenvalue problems. In terms of computational complexity, they are tractable as long as $R$, the number of electron configurations used, grows at most polynomially with the number of qubits for a given accuracy. However, for some strongly correlated systems, the number of significant configurations may scale exponentially with the number of qubits, which can limit the applicability of the method. Nevertheless, the relevant subspace can still be significantly smaller than the entire Fock space in practice. Indeed, QSCI can make it feasible to treat certain strongly correlated systems far exceeding 50 qubits, well beyond the reach of exact diagonalization, as illustrated in Ref.~\cite{kanno2023}.

\section{\label{sec:res}Experiments
}

We now demonstrate the quantum computation for the quasiparticle band structure using a real quantum device. We also show results of supplementary numerical simulations.

\subsection{Settings}

\begin{figure}[t]
\includegraphics[width=7.5cm]{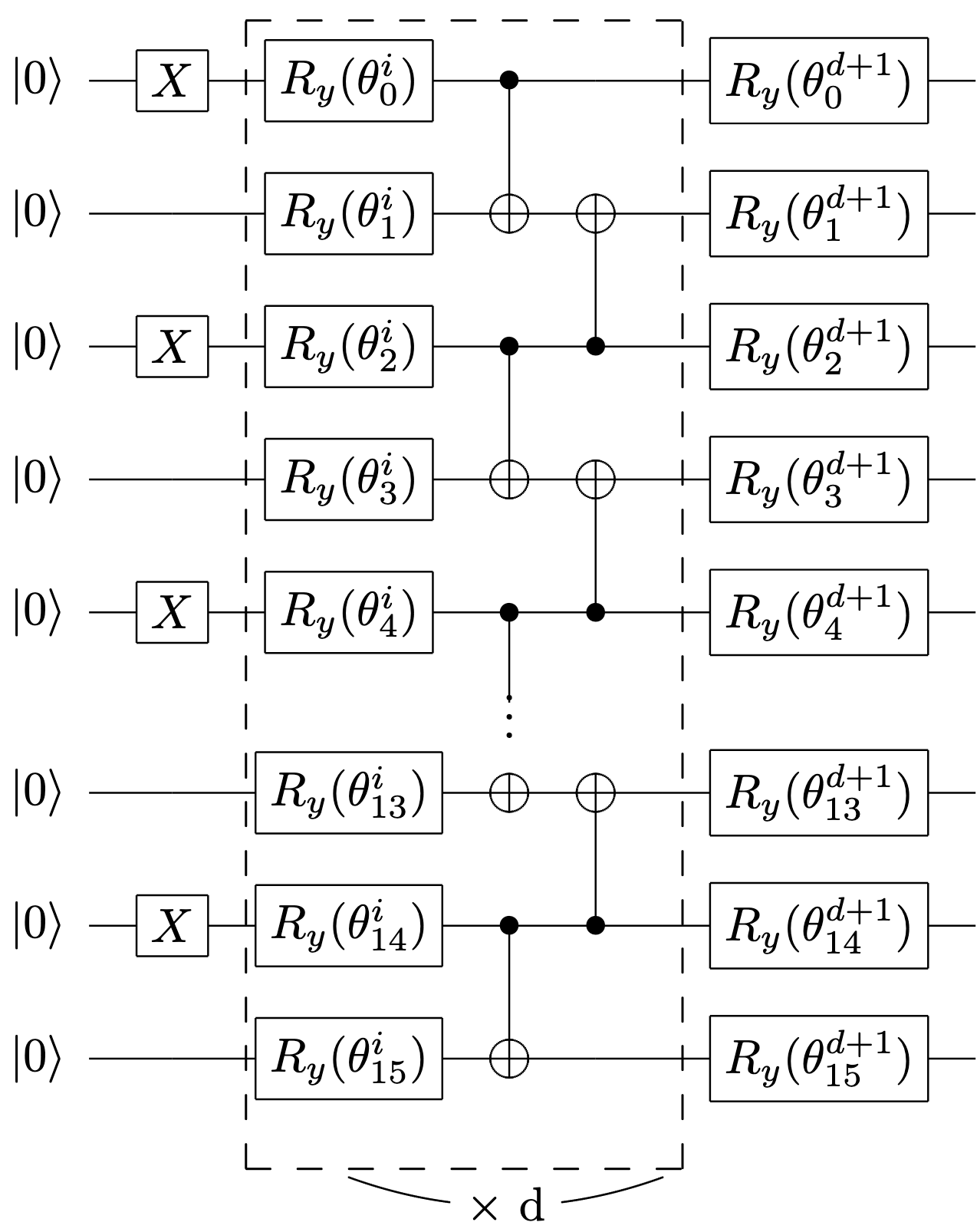}
\caption{\label{fig:ansatz}(Color online) 
Structure of the quantum circuit used as VQE ansatz in this study. The HF state, $\ket{10101\cdots 010}$, is generated by applying $X$ gates to the initialized state $\ket{\psi_0}=\ket{0}^{\otimes 16}$ of 16 qubits. Note that we adopt an unusual ordering in the mapping of spin orbitals to qubits as described in the main text. $R_y(\theta)$ represents the rotation about the $y$-axis. $d$ is the number of repeated layers, or depth. $16(d+1)$ variational parameters of $\theta_{0\mathchar`- 15}^{(i)}$ ($i=1,\cdots, d+1$) are optimized during VQE calculations. A partially optimized state is used as the input state for QSCI. We take $d=3$ in the numerical study.
}
\end{figure}

We apply the QSE-QSCI approach to silicon of diamond crystal structure with the experimental lattice constant 5.43 \AA. For $k$-point sampling of the Brillouin zone, we take a $1\times 1 \times 1$ grid\footnote{It should be noted that extrapolation to the infinite-$k$-point limit is necessary for comparison with experimental results. However, as this would require significantly more qubits, we leave this for future studies.} centered on the target $k$ point along the band path.
The HF calculation and the Hamiltonian construction are carried out separately for each target $k$ point.
With the GTH-SZV (single-zeta valence) basis set and GTH pseudopotential~\cite{goedecker1996}, the Si crystal is described by 8 COs and 8 electrons for each $k$ point. This corresponds to a 16-qubit system after the Jordan-Wigner transformation that we adopt to map the fermionic Hamiltonian Eq.~(\ref{eq:ham}) into the qubit Hamiltonian.
Using the Hamiltonians constructed this way, the QSE-QSCI computation is executed separately for each target $k$ point.
Throughout this work, the second-quantized Hamiltonians are generated by using PySCF\footnote{We encountered a version dependence issue of PySCF in HF calculations. The version 2.3.0 is consistently used in this paper.}~\cite{sun2020} and OpenFermion~\cite{mcclean2020}, with the crystal structure constructed by ASE~\cite{larsen2017atomic}. Numerical simulations are conducted using Qulacs~\cite{suzuki2021} and QURI Parts~\cite{quri}.

Figure~\ref{fig:ansatz} shows the quantum circuit that we use as the VQE ansatz in this study. It is designed based on the hardware-efficient ansatz~\cite{kandala2017}, with several modifications incorporated. First, the circuit is simplified to reduce the number of gates and, consequently, alleviate the effects of noise. Second, each two-qubit gate acts on a pair of qubits corresponding to occupied and unoccupied spin orbitals in the HF state, which is expected to efficiently capture electron correlation with a shallow circuit. Note that occupied and unoccupied spin orbitals are represented by $\ket{1}$ and $\ket{0}$ in the Jordan-Wigner mapping, respectively, and we adopt an unusual orbital ordering for the HF state, $\ket{10101\cdots 010}$, to simplify the ansatz circuit.
In the following, we take $d=3$ for the depth of the ansatz circuit.

\begin{figure}[t]
\includegraphics[width=7.5cm]{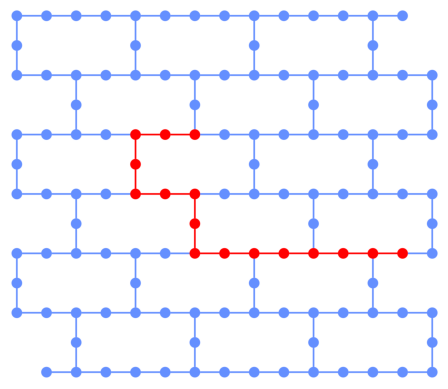}
\caption{\label{fig:qubit_config}(Color online) 
Configuration of the 127 qubits on the IBM Quantum processor \device{kyoto}. The 16 qubits used in this study are highlighted. See the main text for details.
}
\end{figure}

We perform VQE by the noiseless state-vector simulation. For the cost functions to be optimized, we add the following penalty terms~\cite{mcclean2016,ryabinkin2018,kuroiwa2021} to the Hamiltonians: $0.5(\hat{N}_e - 8)^2 +0.2(\hat{S}_z)^2$, where $\hat{N}_e$ and $\hat{S}_z$ are operators for the particle number and the total spin in $z$-direction for electrons, respectively. The terms enforce the resulting states to have $N_e=8$ and $S_z=0$.
The Broyden–Fletcher–Goldfarb–Shanno (BFGS) algorithm is employed to iteratively optimize the variational parameters via a library SciPy~\cite{virtanen2020scipy}.

For moderately optimized ansatz states, we conducted the sampling task by executing a quantum device. In this study, we used \device{kyoto}, which is one of the IBM Quantum Eagle processors~\cite{ibmQC}. We used 16 out of 127 qubits as shown in Fig.~\ref{fig:qubit_config}.
They are high-fidelity qubits arranged on a one-dimensional path, chosen such that the sum of the two-qubit gate error rates is minimized.
For the entire set of 127 qubits, the average error rates for single-qubit gates, two-qubit gates, and readout were 0.3\%, 4.2\%, and 3.1\%, respectively, while they were improved to 0.02\%, 0.5\%, and 1.4\% for the selected 16 qubits.
The acquisition of these data and the execution of quantum circuits were carried out on December 22, 2023 (JST).
We used Qiskit~\cite{qiskit2024} for the circuit execution as well as the optimization of the circuit to reduce its depth and gate count by the transpiler with the optimization level of 3; no error mitigation was applied at this stage.
We took the VQE state at the iteration step of $400$ and repeated the circuit execution  $N_{\rm shot}=10^4$ times for each of representative $k$ points on the band path. The sampling outcomes are post-selected according to the conserved quantities of the particle number and spin $S_z$ for electrons. Then, $R$ configurations, or Slater determinants, are selected to construct the subspace for QSCI.

For comparison, we also perform noiseless simulations for the sampling task in QSCI.
They are based on the ideal sampling method~\cite{kanno2023}, where $R$ configurations having the largest amplitudes in the input-state wave function are selected to define the subspace $\mathcal{S}_R$.

\subsection{Quasiparticle Band Structure}

\begin{figure}[t]
\includegraphics[width=8.cm]{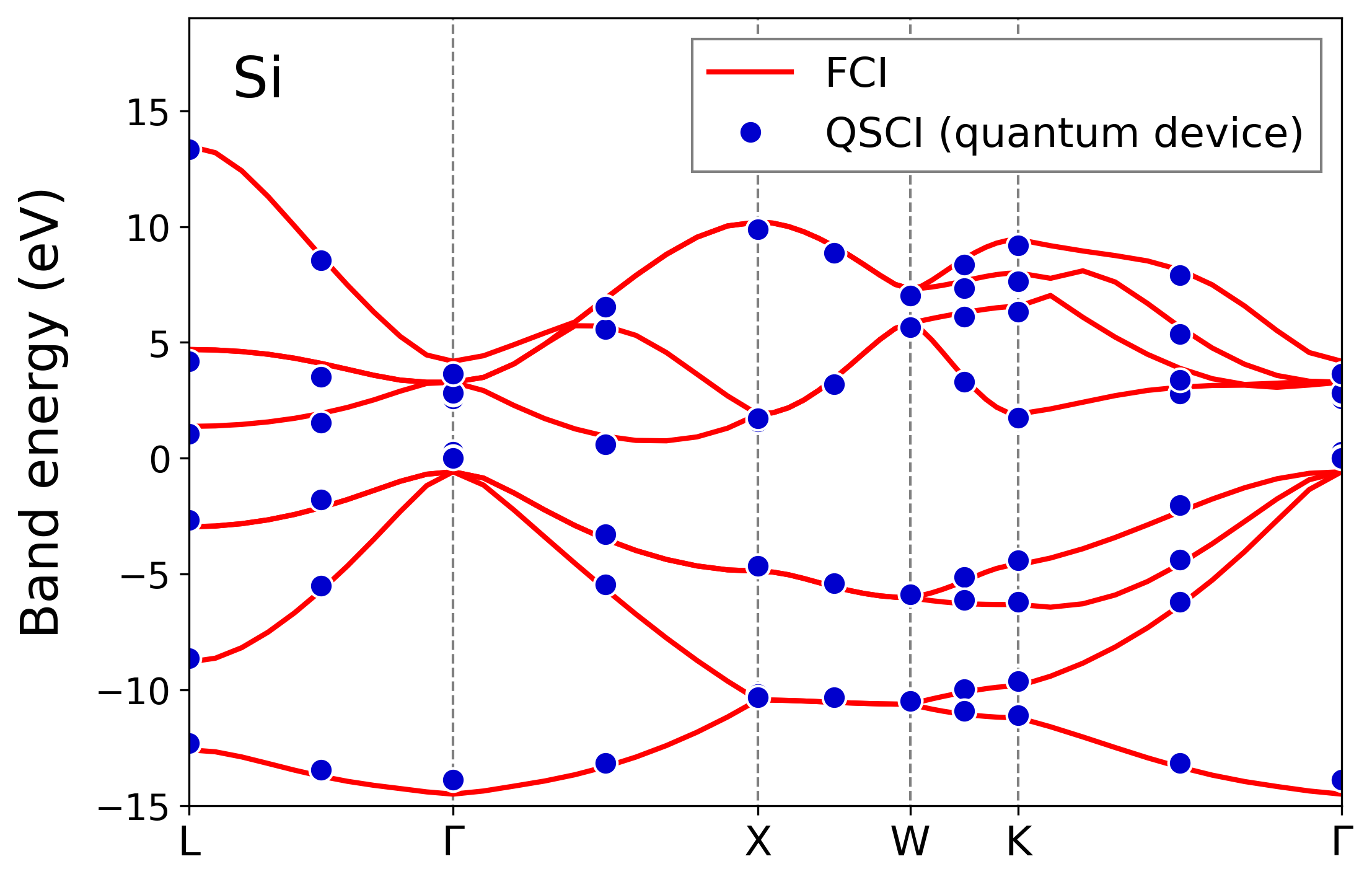}
\caption{\label{fig:band}(Color online) 
Quasiparticle band structure of the Si crystal obtained by our QSE-QSCI approach with 16 qubits of the quantum device \device{kyoto}. 50 most frequent configurations are used in the QSCI calculation. For comparison, we also show the result based on FCI for the ground-state calculation combined with QSE.
}
\end{figure}

Figure~\ref{fig:band} shows the quantum-device result for the quasiparticle band structure obtained by the QSE-QSCI approach. In the QSCI calculation, $R=50$ configurations passing the post-selection are included. We also put the exact result, where the QSCI calculation is replaced by the exact diagonalization, or FCI, for comparison. We observe that the quantum-device calculation shows good agreement with the FCI result.
Further details are provided in the following subsections.

%

\subsection{Ground States}

\begin{figure*}
    \includegraphics[width=\textwidth]{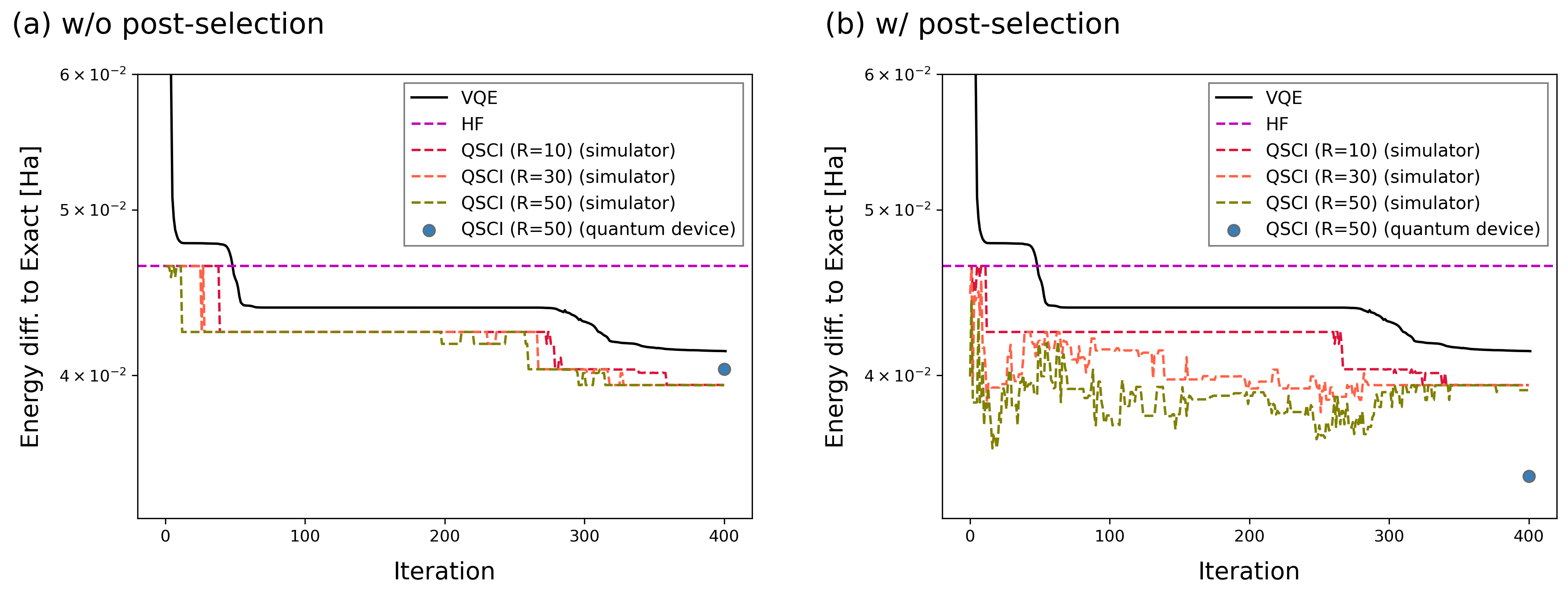}
    \caption{(Color online) QSCI results for the ground state of the Si crystal, shown with the VQE optimization history at the $L$ point on the band path. The QSCI results are obtained by the noiseless simulation or the quantum device with the sampling outcomes (a) before or (b) after the post-selection.
    In the noiseless simulation, each of the VQE states are used as the input for QSCI with the number of electron configurations varied ($R = 10, 30, 50$). The quantum-device result of QSCI is shown for the 400th iteration step.
    The energy obtained by each method is plotted as the difference from the FCI energy in Hartree. The energy obtained by the HF calculation is also shown for reference.
    }
    \label{fig:optimization}
\end{figure*}

In Fig.~\ref{fig:optimization}, we show the optimization history of VQE where the $L$ point on the band path is chosen for illustration. The results of QSCI were obtained from the noiseless simulation, where the VQE state at each iteration step is used as the input state.
The results based on the quantum device are also shown at the 400th iteration step.
The panel (a) shows QSCI results based on the raw sampling outcomes, i.e., without post-selection, for different $R$ values. By contrast, the QSCI results shown in the panel (b) rely on the post-selected sampling outcomes.

We observe that the QSCI result tends to improve as the VQE state is further optimized.
We also see that the post-selection is generally effective to improve the QSCI energies. In particular, the result obtained from the quantum device is even worse than the HF calculation without post-selection, while it becomes consistent with the noiseless simulation result after post-selection.

Note that the noiseless simulation results are also affected by post-selection, which may be somewhat counterintuitive. This is because the VQE ansatz we adopt is not explicitly designed to conserve the particle number and spin. Although the penalty terms are added to the cost function in VQE, they are not restrictive enough to completely prevent contamination from sectors with different values of the conserved quantities.
Nevertheless, QSCI with post-selection can effectively eliminate such contamination.

%

\subsection{Distribution of Electron Configurations}

\begin{figure}[t]
\includegraphics[width=8.5cm]{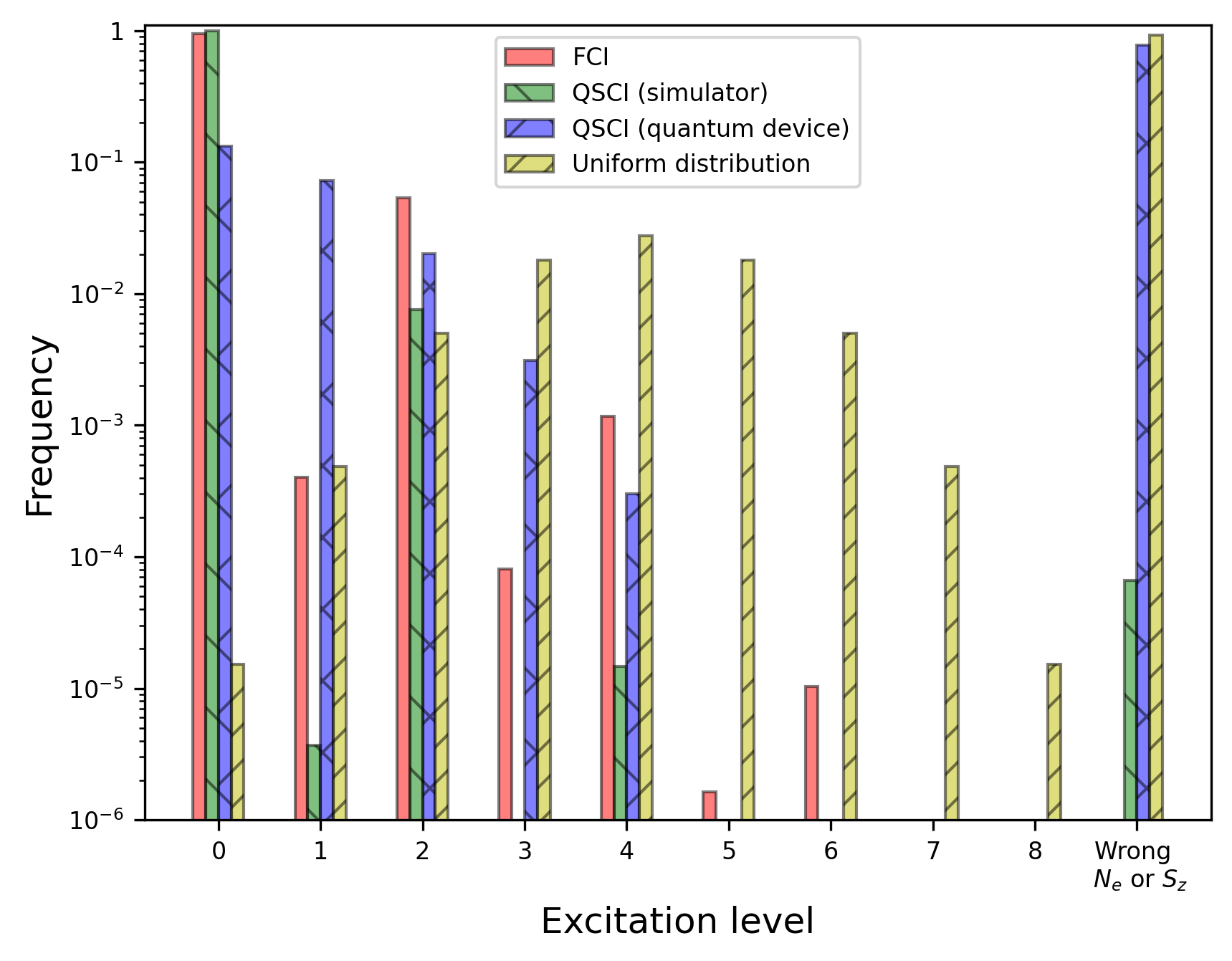}
\caption{\label{fig:distribution}(Color online) 
Distributions of electron configurations in sampling outcomes for the $L$ point on the band path. Configurations are categorized according to the excitation level in the Slater determinants, and whether the particle number $N_e$ and spin $S_z$ are conserved or not. The results are shown for the ideal sampling of the FCI wave function, the noiseless simulation and quantum-device results of QSCI for the VQE state at the 400th iteration step, and the uniform distribution. See the main text for further details.
}
\end{figure}

Figure~\ref{fig:distribution} shows the distributions of sampling outcomes for both the noiseless simulation and the quantum device. The electron configurations are classified according to the excitation level: for instance, level 1 corresponds to single excitations, and level 2 corresponds to double excitations. The frequency of each excitation level in the sampling outcomes is then plotted.
The frequency of configurations having a wrong particle number or spin is also shown for each case.

The uniform distribution (with respect to the electron configurations) is also shown. This corresponds to the fully-depolarized case owing to noise, where the quantum state is described by a mixed state $\rho \propto I$.
We observe that the distribution obtained from the quantum device reasonably captures the excitations present in the ideal noiseless simulation, while retaining the shape distinct from the worst case of full depolarization.

The distribution corresponding to the FCI wave function is also shown for reference.
We see that, unlike in FCI, configurations with excitation levels 5 or 6 are absent (at this precision) even in the noiseless simulation. This suggests that further improvements in state preparation would lead to an improvement in the QSCI result.

To further assess the quality of sampling results from the real quantum device, we employ a statistical distance metric, namely, the Jensen-Shannon (JS) divergence~\cite{lin1991divergence}, which satisfies $0 \leq D_{\rm JS} \leq 1$. (See Appendix~\ref{app:JS} for details.) We use the JS divergence as a measure to evaluate how closely the distribution obtained from sampling on the quantum device resembles either the ideal noiseless distribution or the noisy uniform distribution.
The JS divergences calculated for pairs of distributions in Fig.~\ref{fig:distribution} are $D_{\rm JS} = 0.25$ between the quantum device and simulator, while $D_{\rm JS} = 0.95$ between the quantum device and uniform distribution.
Thus, we quantitatively confirm that the sampling distribution obtained from the quantum device is significantly closer to the ideal noiseless distribution than to the noisy uniform distribution, supporting the reliability of the quantum-device results.

%

\section{Summary and Outlook}
In this paper, we demonstrated a hybrid quantum-classical approach for calculating a quasiparticle band structure using QSE combined with QSCI. QSCI was introduced as an alternative to VQE for ground-state calculations. It utilizes quantum computers to sample relevant electron configurations and classical computers for subspace Hamiltonian construction and diagonalization. Unlike VQE, QSCI does not require full optimization of variational parameters of the input state, making it more robust against device noise and statistical errors. By combining QSCI with QSE, we were able to efficiently compute a quasiparticle band structure. Specifically, we successfully demonstrated the feasibility of our approach by calculating the quasiparticle band structure of a silicon crystal using 16 qubits on an IBM quantum processor. The results showed good agreement with exact diagonalization (FCI) calculations, validating the effectiveness of our method.

In the present study, we still used VQE as the input state preparation for QSCI, and thus, the problems on the parameter optimization for large systems are not completely resolved. To address this issue, promising alternatives include adaptive construction of the input state for QSCI (ADAPT-QSCI) ~\cite{nakagawa2024} or the use of time evolution to prepare the input states~\cite{sugisaki2024, mikkelsen2024quantum, yu2025quantum}. In addition, the use of the local unitary cluster Jastrow ansatz~\cite{matsuzawa2020jastrow,motta2023bridging} with variational parameters coming from a Coupled-Cluster Singles and Doubles (CCSD) calculation is another way to bypass the need for a costly variational optimization~\cite{robledo2024chemistry,duriez2025computingbandgapsperiodic}. Applying these methods to quasiparticle band structures would be intriguing subsequent studies. 




\section*{Acknowledgements}

We thank Shih-Yen Tseng for valuable contributions during the initial phase of the study. H. I. and M. K. also thank Kentaro Wada for assistance in circuit transpilation.
We acknowledge the use of IBM Quantum services for this work. The views expressed are those of the authors, and do not reflect the official policy or position of IBM or the IBM Quantum team.

\appendix

\section{Jensen-Shannon Divergence \label{app:JS}}

Here we give the definition of the Jensen-Shannon (JS) divergence~\cite{lin1991divergence} with some of basic properties.

For a pair of probability distributions $P(x)$ and $Q(x)$, the Kullback-Leibler divergence is defined by
\begin{align}
    D_{\rm KL}(P \parallel Q)
    = \sum_{x} P(x) \log \left( \frac{P(x)}{Q(x)} \right).
\end{align}
Then, the JS divergence between $P(x)$ and $Q(x)$ is defined by
\begin{align}
    &D_{\rm JS}(P \parallel Q)\notag\\ 
    &= \frac{1}{2}D_{\rm KL}\left( P \parallel \frac{P+Q}{2} \right) 
    +\frac{1}{2}D_{\rm KL}\left( Q \parallel \frac{P+Q}{2} \right).
\end{align}
The JS divergence is a symmetrized and bounded version of the KL divergence satisfying the following propeties:
\begin{align}
&D_{\rm JS}(P \parallel Q) = D_{\rm JS}(Q \parallel P), \\
&0 \leq D_{\rm JS}(P\parallel Q) \leq \log 2.
\end{align}
$D_{\rm JS}(P\parallel Q)=0$ if and only if two distributions are identical, i.e., $P(x)=Q(x)$ for all $x$.

In this paper, we adopt the JS divergence as a measure to evaluate how closely a distribution obtained from sampling on the noisy quantum device resembles the ideal noiseless distribution.
We define the logarithm with base 2; hence, the upper bound of the inequality is 1, i.e., $0 \leq D_{\rm JS}(P\parallel Q) \leq 1$.

\bibliography{reference.bib}

\end{document}